\documentstyle[12pt]{article}
\def\be{\begin{equation}}
\def\ee{\end{equation}}
\def\d{\partial}
\def\rref#1{(\ref{#1})}
\def\df#1#2{\frac{\d #1}{\d #2}}
\def\bra{\langle}
\def\ket{\rangle}
\def\cH{{\cal H}}
\def\ie{i.e.\,}
\begin{document}
\title{Wavelet basis for the Schr\"{o}dinger equation}
\author{M.V.Altaiski\thanks{Permanent address: Joint Institute
for Nuclear Research, Dubna, 141980, RUSSIA.} \\
Centre for Applicable Mathematics,
B.M.Birla Science Centre, \\
Adarshnagar, Hyderabad, 500463, India}
\date{}
\maketitle
\begin{abstract}
The self-similar representation for the
Schr\"{o}dinger equation \\ is derived.
\end{abstract}

The general form Schr\"{o}dinger equation
\be
\imath\hbar \df{\psi}{t} = \hat H \psi, \label{se}
\ee
which is the generalization of the free particle wave equation
$$ \imath\hbar \df{\psi}{t} =
-\frac{\hbar^2}{2m}\frac{\d^2\psi}{\d x^2},
$$
describes the evolution of the wave function $\psi(\vec x,t)$ for
the quantum system with a general form of Hamiltonian $\hat H$. The solution
of the Schr\"{o}dinger equation \rref{se}, if it exists, is not
uniquely determined by the Hamiltonian; the symmetry of the
problem and/or boundary conditions must be given also.

For this reason, for the problems with  spherical symmetry ($SO_3$)
--- say, the Hydrogen atom --- we look for a solution within the
class of spherical functions. For homogeneous problems, \ie the
problems invariant under translations, we suppose the solution to
be the superposition of plane waves, the unitary representations of
the translation group
\be
G: x' = x+b. \label{tr}
\ee
To find a solution of the Schr\"{o}dinger equation
\rref{se} with a (Lie) symmetry group $G$, we have to look for a
state vector $|\psi\ket$ of a Hilbert space $\cH$ in the form
with respect to $G$
\be
|\psi\ket = c_v \int_G U(g)|v\ket d\mu_L(g)\bra v|U^*(g)|\psi\ket.
\label{du}
\ee
This means that we suppose it to be a decomposition with respect to
a certain representation $U(g)$ of the Lie group $G$; $d\mu_L(g)$ is the
left-invariant measure on $G$. Here after $v\in\cH$
an {\em admissible} vector of the representation $U(g)$ \cite{gr,av},
a vector of the Hilbert space $\cH$ satisfying the normalization
condition
\be
c_v = \int_G | \bra v | U(g)|v\ket|^2 d\mu_L(g) < \infty
\ee
is referred to as a basic wavelet \cite{daub}.
For the group of translations \rref{tr} the decomposition
\rref{du} is a trivial one
$|\psi\ket = \int |k\ket dk\bra k|\psi\ket$
and causes no problem with the particular choice of the admissible
vector. For other groups this problem is not always trivial and
often requires some physical insight.

In the present paper, following \cite{ss}, we consider the possibility
of self-similar objects in quantum mechanics and related solutions
of the Schr\"{o}dinger equation.

For the case of the one dimensional  Schr\"{o}dinger equation
\be
\imath\hbar \df{\psi}{t} = \left[{
-\frac{\hbar^2}{2m}\frac{\d^2}{\d x^2}+W(x)
}\right] \psi, \label{se1}
\ee
(taken for simplicity) the Lie group which pertains to the
self-similarity is the affine group
\be
G: x'=ax+b; \label{ag}  \qquad d\mu(a,b) = \frac{dadb}{a^2}
\ee
The representation of the affine group \rref{ag} apt to the problem
ultimately has the form
\be
U(a,b)f(x) = \frac{1}{\sqrt{a}}
f\left(\frac{x-b}{a}\right),
\ee
but the particular type of the basic wavelet $v(x)$ is to be
chosen from physical consideration.

For the sake of a well defined quasiclassical limit at, large $a$
we have to choose a minimal wave packet
\be
v_0(x) = [2\pi(\Delta x)^2]^{-1/4} \exp \left({
-\frac{(x-\bar x)^2}{4(\Delta x)^2} + \imath p x }\right)
\label{mp}
\ee
as a basic wavelet; since it minimises the uncertainty relation
$\Delta x \Delta p \ge \hbar/2$ (See e.g. \cite{shiff}).
Strictly speaking, the minimal wave packet \rref{mp} is not an
admissible vector, since $c_{v_0}$ is formally equal to infinity.
This ambigity, however, is not of principal importance, and in
our model we can treat $c_{v_0}$ as a formal constant.

Substituting,
\be \psi(x) = c_v^{-1} \int \frac{1}{\sqrt{a}}
v\left(\frac{x-b}{a}\right)\bra v;a,b|\psi \ket \frac{dadb}{a^2}
\ee
into the Schr\"{o}dinger equation \rref{se} we get
\begin{eqnarray}
\imath\hbar\frac{\d}{\d t} \bra v; a,b|\psi\ket = \\
\nonumber \frac{\hbar^2}{2ma^2}
\left[{
\frac{1}{(\Delta x)^2} -
\left\{ {
\frac{\imath p}{\hbar} - \frac{\frac{x-b}{a}-\bar x}{(\Delta x)^2}
}\right\}
}\right]\bra v;a,b|\psi \ket
 + W(a,b)\bra v;a,b|\psi \ket
\end{eqnarray}
where $W(a,b)$ is the wavelet transform of the potential $W(x)$:
$$
W(a,b) = c_v^{-1} \int \frac{1}{\sqrt a} v\left(\frac{b-x}{a}\right)
W(x)dx
$$
This potential, in our opinion, might have a deep physical
meaning, since the potential function, which describes the
interaction at quantum level, will ultimately depend on scale,
rather than only distance \cite{not}; this is exactly the case
for $W(a,b)$.

Near the central point $\frac{x-b}{a}\approx\bar x$, \ie close to
the "actual" position of the particle, we have
\be
\imath\hbar\df{}{t} \bra v; a,b|\psi \ket \approx
\nonumber \Bigl[
\frac{p^2}{2ma^2} +  \frac{\hbar^2}{4ma^2(\Delta x)^2}
+ W(a,b) \Bigr]
\bra v;a,b|\psi \ket
\ee

\end{document}